\documentclass{ws-procs9x6}
\usepackage{epsfig}
\usepackage{colordvi}
\usepackage{latexsym}
\usepackage{bm}

\def\draftnote{}
\def\trimmarks{}

\begin{document}

\title{EQUILIBRIUM THERMODYNAMICS OF LATTICE QCD}

\author{D.~K.~Sinclair}

\address{HEP Division, Argonne National Laboratory, 9700 South Cass Avenue,
Argonne, IL, 60439, USA}

\begin{abstract}
Lattice QCD allows us to simulate QCD at non-zero temperature and/or densities.
Such equilibrium thermodynamics calculations are relevant to the physics of
relativistic heavy-ion collisions. I give a brief review of the field with
emphasis on our work.
\end{abstract}

\bodymatter 












\section{Introduction}

High temperature hadronic matter was certainly present in the early universe.
Relativistic heavy-ion colliders such as RHIC and in future the LHC with
heavy-ions, produce hot hadronic matter. Lower energy machines can produce
hot hadronic matter with an appreciable baryon/quark-number density -- hot
nuclear matter. At high enough temperatures this hadronic matter is expected
to become quark/gluon matter -- the quark-gluon plasma. Preliminary evidence
for the quark-gluon plasma has been reported from RHIC and CERN.

At high baryon/quark-number density and low temperature such as might exist in
the cores of neutron stars, exotic states of matter, such as colour
superconducting phases have been proposed.

In Quantum Chromodynamics (QCD), the accepted theory of hadrons and their
strong interactions, the most interesting finite-temperature properties are
non-perturbative. Lattice QCD simulations are the only systematic approach to 
the study of such phenomena. 

\begin{figure}[htb]
\epsfxsize=3.5in
\centerline{\epsffile{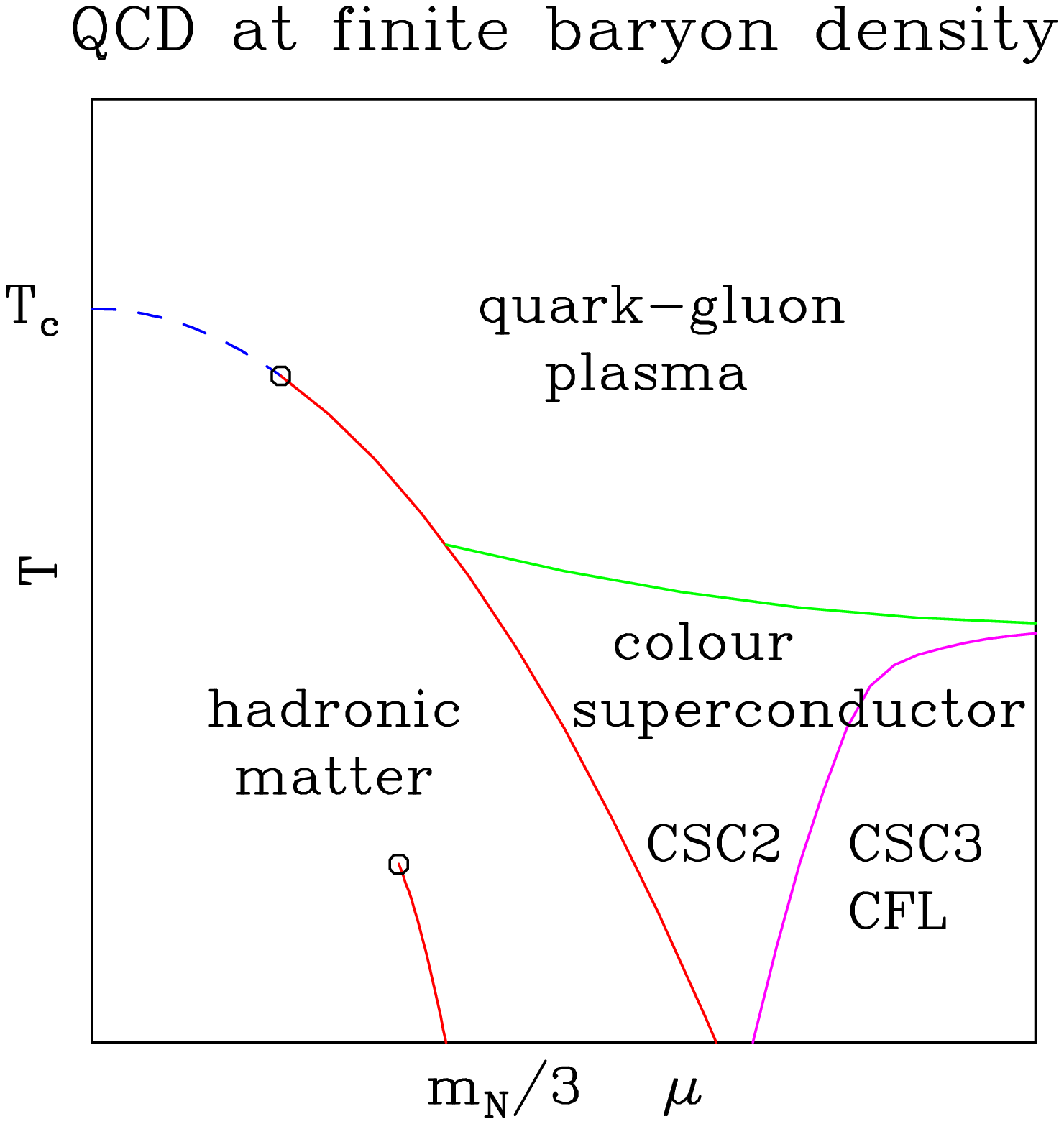}}
\caption{Simplified phase diagram for QCD at finite temperature and 
quark-number chemical potential $\mu$.}
\label{fig:phase}
\end{figure}

Fig. \ref{fig:phase} shows a simplified version of the proposed phase diagram.
(For an introduction to the phase structure of QCD and predictions from lattice
QCD, see recent reviews \cite{Heller:2006ub,Alford:2001dt}).

In Lattice QCD, we only really know how to calculate equilibrium thermodynamics.
Relativistic heavy-ion collisions are clearly not equilibrium processes, but
knowledge of equilibrium thermodynamics and the consequent extraction of
thermodynamic quantities and the prediction of the equation-of-state give us
some idea of the type of behaviour we might expect. In addition, it can provide
input for hydrodynamic models of such collisions.

QCD at finite temperature and/or densities yields information on the dynamics
of confinement and chiral symmetry breaking and thus enhances our understanding
of QCD.

\section{Lattice QCD at finite temperature}

QCD is quantized by functional integral methods. To make these integrals
well-defined, we rotate to imaginary time -- Euclidean space-time. The quark
fields are then defined on the sites of a 4-dimensional hypercubic lattice,
the gauge fields $U_\mu = \exp(i g A_\mu)$ on its links.
\begin{equation}
Z = \int {\cal D}\psi{\cal D}\bar{\psi}{\cal D}U e^{-S}
\end{equation}
where the action $S=S_f+S_g$.
The simplest lattice implementation (which preserves gauge invariance) has
\begin{equation}
S_g = \beta \sum_\Box \left( 1- \frac{1}{3}{\rm Re Tr}_\Box UUUU \right) ,
\end{equation}
where $\beta=6/g^2$, and $S_f = \sum_{sites} \bar{\psi}( D\!\!\!\!/ + m ) \psi$.

Simulations are performed by replacing the fermion fields with boson fields
(pseudo-fermions) which results in replacing the Dirac operator by its inverse.
Adding a `kinetic' term for the gauge fields, allowing them to evolve in a
fictitious `time', turns $Z$ into the partition function for a system of
classical particles. One then uses molecular-dynamics techniques to effectively
`evaluate' the integrals, by numerically integrating the equations of motion
for this system. Bringing the system in contact with a heat-bath at regular
intervals assures ergodicity and compensates for the lack of dynamics for the
pseudo-fermion fields.

To compensate for the doubling problems for the fermion fields frequently 
requires taking fractional powers of the fermion determinant. This is performed
by using rational approximations to fractional powers of the Dirac operator.
Use of a global Metropolis accept/reject step removes the discretization errors
in the numerical integration of the equations of motion. (For a recent
description of simulation methods with dynamical fermions see 
\cite{Kennedy:2006ax}).

If we use a lattice of temporal extent $1/T$ and a spatial extent much larger
than this, and demand periodic boundary conditions for the gauge fields and
antiperiodic boundary conditions for the fermion fields in the time direction,
$Z=Z(T)$ the partition function for lattice QCD at temperature $T$.

\section{The finite temperature transition}

For massless quarks, chiral symmetry is restored at the finite temperature
transition. $\langle\bar{\psi}\psi\rangle \ne 0$ below the transition and
$\langle\bar{\psi}\psi\rangle = 0$ above the transition. Hence this transition
is a phase transition. Arguments based on dimensional reduction predict that
this phase transition is a second order phase transition (critical point) for
$N_f=2$, in the universality class of the 3-dimensional $O(4)$ spin model
\cite{Pisarski:1983ms}. 
For non-zero quark mass it is expected to weaken to a crossover with
no real phase transition. For $N_f > 2$ this transition is predicted to be
a first-order phase transition which is expected to remain first-order for
small quark masses, becoming a crossover for larger quark masses.

\begin{figure}[htb]
\epsfxsize=4in
\centerline{\epsffile{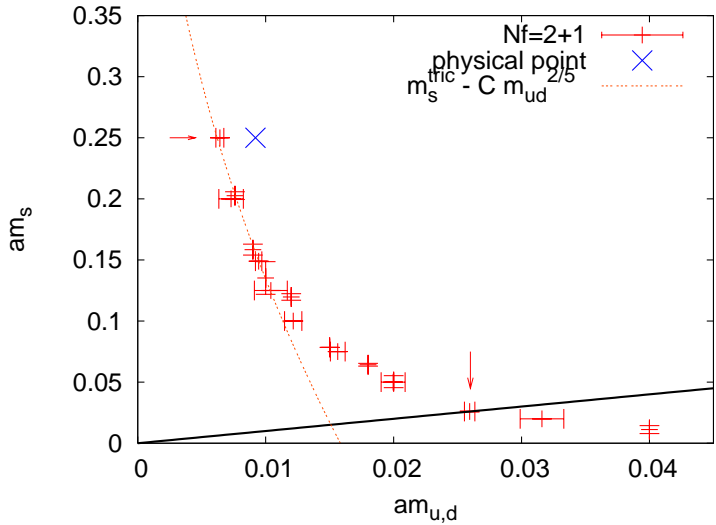}}  
\caption{The chiral critical line at zero baryon-number density (from
Ref.~\cite{deForcrand:2006pv}).}
\label{fig:deF&P}
\end{figure}

In the real world where $N_f=2+1$ ($u,d,s$), the important question is whether
the strange quark mass is light enough for the transition to be first-order or
whether it is a crossover. Recent lattice simulations indicate that the strange
quark mass is too large, and the transition is a crossover 
\cite{deForcrand:2006pv} (see Fig~\ref{fig:deF&P}). This extends earlier work
of \cite{Karsch:2001nf}
(NB the $N_f=3$ point agrees with our own simulations.) 

Simulations with larger lattices, finer lattice spacings and improved actions
have recently been reported by \cite{Aoki:2006we}.
Here the $u$, $d$ and $s$ quark masses are chosen by fixing $m_K/m_\pi$ (and
$f_K/f_\pi$) at their physical values. Making several choices of masses 
restricted in this manner, they were able to extrapolate to the physical quark
masses and below. They confirm the observation that the physical transition  
is a crossover and not a phase transition, in simulations which probe much 
closer to the continuum limit than in \cite{deForcrand:2006pv}. The same
simulations give new estimates for the temperature(s) of this crossover
\cite{Aoki:2006br}.

Since the strange quark appears too massive to control the nature of the 
transition, it is useful to study the 2-flavour case, to see if the nature of 
the $m=0$ phase transition agrees with the above prediction. 

Determining the nature of the 2-flavour phase transition has proved difficult,
since is expected only at $m=0$, and finite volume effects appear to be large
\cite{Karsch:1993tv,Karsch:1994hm,Aoki:1998wg,Bernard:1996cs}.
Standard (including highly improved) lattice actions forbid simulations at
$m=0$, and small mass simulations are very expensive. No such simulations have
been performed at masses small enough to uncover the $m=0$ results.

\begin{figure}[htb]
\epsfxsize=3.5in
\centerline{\epsffile{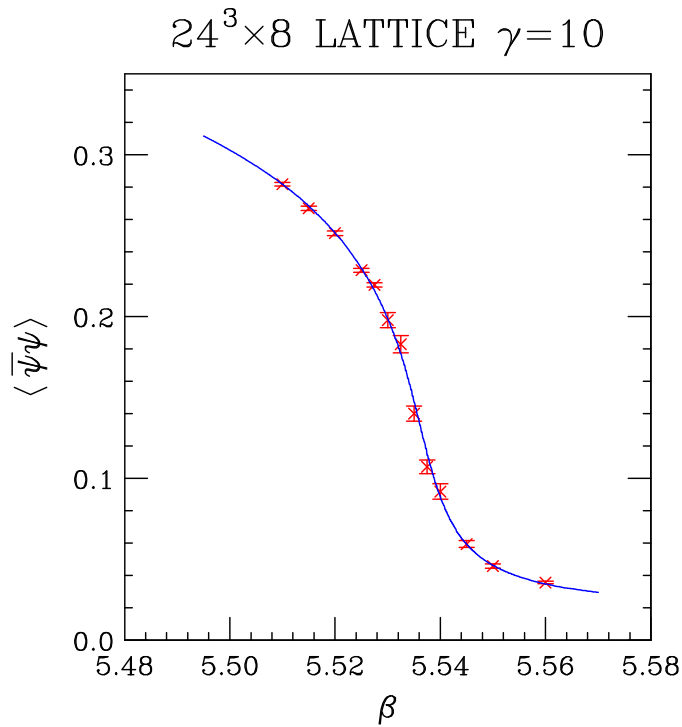}}
\caption{Chiral condensate (points) near the chiral phase transition for
lattice QCD, fitted to the magnetization (solid line) for the $O(2)$ spin 
model (from Ref.~\cite{Kogut:2006gt}).}
\label{fig:O2}
\end{figure}

We have performed $N_f=2$ simulations using the $\chi$QCD action which allows 
simulations at zero quark mass \cite{Kogut:2006gt}. 
Since it is a staggered quark action, the
reduced flavour symmetry leads us to predict that the phase transition should 
be second-order and in the universality class of the 3-dimensional $O(2)$
spin model. To accommodate the finite volume effects, we compare our chiral
condensate measurements to the magnetization of the $O(2)$ spin model also on
finite volumes, as shown in Fig.~\ref{fig:O2}. 
The agreement is excellent. Good agreement is also obtained
with correlation lengths and susceptibilities. 

\begin{figure}[htb]
\epsfxsize=4in
\centerline{\epsffile{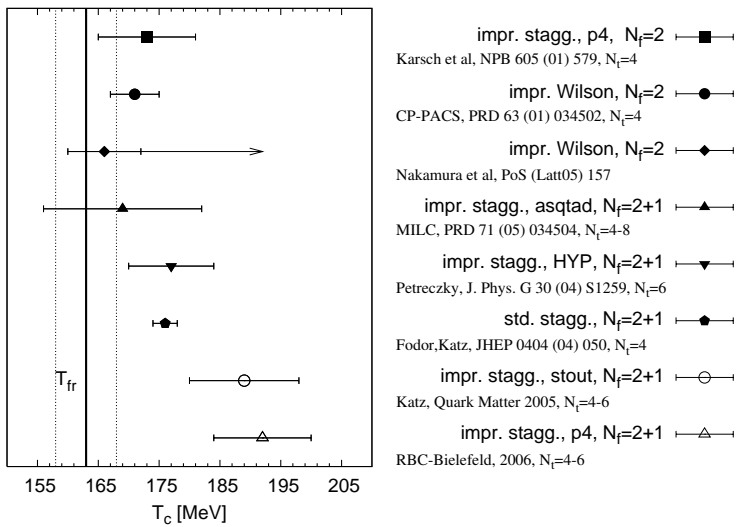}}
\caption{Recent lattice estimates of $T_c$ compared with experiment
(from Ref.~\cite{Petreczky:2006zk}).}
\label{fig:Petreczky}
\end{figure}

Fig.~\ref{fig:Petreczky} shows recent estimates of the transition temperature
$T_c$ from lattice QCD simulations, compared with an experimental estimate
from RHIC. All lattice simulations are consistent with the chiral and
deconfinement transitions being coincident.

\section{The equation-of-state for hot QCD.}

The equation-of-state (EOS) expresses the pressure $p$, the entropy density $s$
and the energy density $\epsilon$ as functions of the temperature $T$. The
free energy density, pressure and entropy are given by 
\begin{equation}
f = -{T \over V} \ln Z(T) , \;\;\; p = -f \;\;\;, s = {d p \over d T}, 
\end{equation}
respectively. The energy density is not an independent quantity but is given by
\begin{equation}
\epsilon = T s - p.
\end{equation}
$Z(T)$ is calculated on the lattice by numerically integrating
\begin{equation}
{d \ln Z \over d \beta} = \langle S_g \rangle .
\end{equation}
To obtain $T$ as a function of $\beta$ requires knowing the lattice spacing as
a function of $\beta$. This can be obtained by measuring physical quantities
as functions of $\beta$ at zero temperature.

\begin{figure}[htb]
\epsfxsize=4in
\centerline{\epsffile{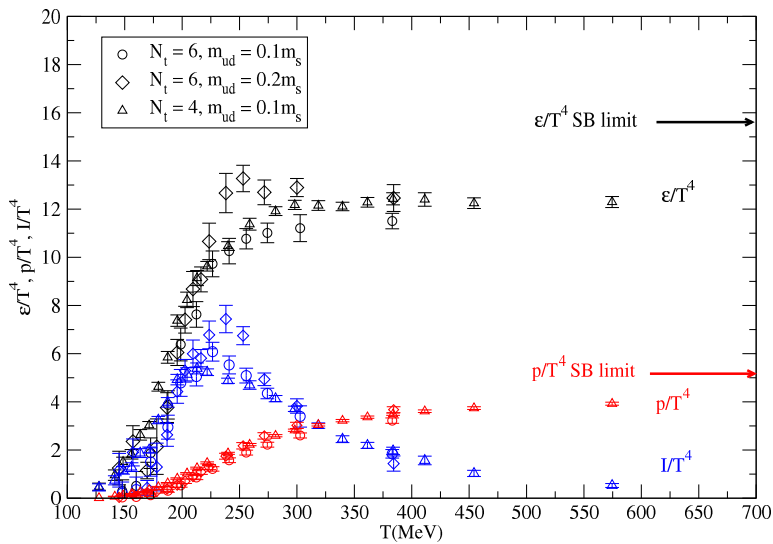}}
\caption{The EOS from Lattice QCD simulations (MILC preliminary: status 
Lattice2006 \cite{Bernard:2006xg}). $I = \epsilon - 3 p$}
\label{fig:MILC}
\end{figure}

Knowledge of the EOS is needed as input to models for the evolution of the
hot hadronic matter in relativistic heavy-ion collisions. Fig.~\ref{fig:MILC} is
a graph of these quantities from the MILC collaboration \cite{Bernard:2006nj}. 

For earlier work on the finite temperature EOS see for example 
\cite{Karsch:2000ps}.
Karsch {\it et al.} have shown that the low temperature behaviour is well
modeled as a non-interacting gas of hadron resonances \cite{Karsch:2003vd}.
See also \cite{Aoki:2005vt} for other work on the QCD EOS.

\section{Meson spectral functions at finite T}

Meson spectral functions yield information about the propagation of mesons, or
excitations with mesonic quantum numbers in hot hadronic matter 
\cite{Petreczky:2004xs} (This review gives references to earlier works). 
Even just
above $T_c$, the quark-gluon plasma is a strongly interacting fluid and mesonic
states survive. Not only do the spectral functions have information about
hadronic stability at high temperatures, but they also have information about
transport coefficients and dilepton production. At zero temperature, the
spectral function is just the momentum space propagator. The Euclidean-time
meson propagator $G(\tau, {\bm p}, T)$ is related to the spectral function
$\sigma(\omega, {\bm p}, T)$ by
\begin{equation}
G(\tau, {\bm p}, T) = \int_{\!\!0}^\infty d\omega 
                      \sigma(\omega, {\bm p}, T) K(\tau, \omega, T)
\end{equation}
where $K(\tau, \omega, T) = {\cosh[\omega(\tau-1/2T)] \over \sinh(\omega/2T)}$.

The main difficulty has been that, since the spatial extent of the lattice
must be much greater than its temporal extent, the temporal extent of the
lattice is typically $\sim 10$ or less in lattice units. This gives a rather 
poor estimate of $\sigma$. Recently simulations have been performed on 
anisotropic lattices on which the spatial lattice spacing is much greater than 
the temporal lattice spacing \cite{Jakovac:2006sf,Aarts:2006nr}. 
This allows for many more points in the time direction, while still
keeping the spatial extent of the lattice much greater than its temporal extent.
This leads to better estimates for the spectral functions.

The charmonium spectral functions have been of particular interest. It had
been suggested that one signal for the quark-gluon plasma phase could be the
`melting' of charmonium. This would reveal itself as a drop in the charmonium
production relative to the production of $D\bar{D}$ pairs and similar states.

\begin{figure}[htb]
\epsfxsize=4in
\centerline{\epsffile{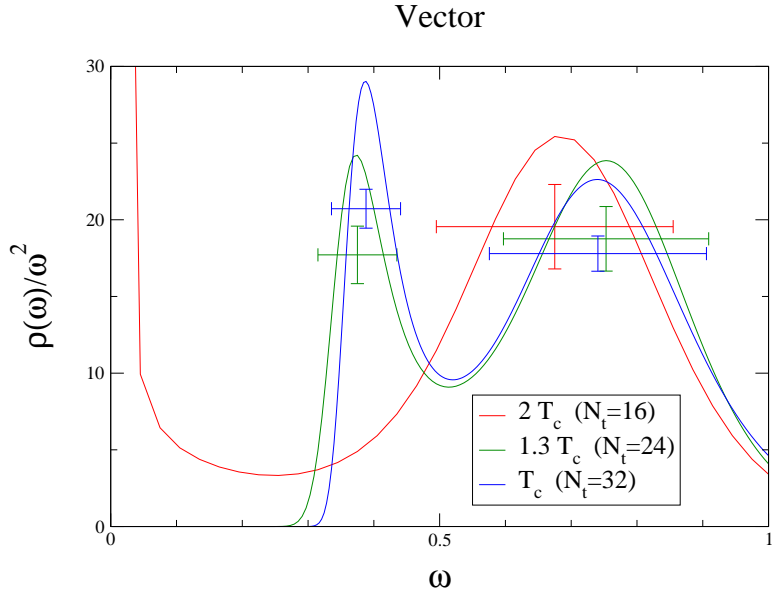}}
\caption{Spectral function $\rho \equiv \sigma$ for the $J/\psi$ at $T_c$,
$1.3 T_c$ and $2 T_c$ (from Ref.~\cite{Aarts:2006nr}).}
\label{fig:trinlat}
\end{figure}

Current simulations by the TrinLat collaboration using anisotropic lattices
with dynamical light quarks, indicate that the $J/\psi$
and $\eta_c$ survive above $T_c$ and melt somewhere between $1.3T_c$ and $2T_c$.
The $\chi_c$ states appear to melt at or below $1.3T_c$ (see 
Fig.~\ref{fig:trinlat}).

At the LHC, even higher temperatures are expected, and it has been suggested
that these might be high enough to melt bottomonia \cite{Gunion:1996qc}. 
We plan to measure
NRQCD bottomonium propagators (and hence spectral functions) on the TrinLat
configurations to determine these melting temperatures.

\section{Transport coefficients}

To use hydrodynamic models for hadronic matter in relativistic heavy-ion
collisions requires knowledge of the transport coefficients, shear viscosity 
$\eta$, volume viscosity $\zeta$ and thermal conductivity $\kappa$. These
viscosities are expressed in terms of Green's functions of the stress-energy 
tensor \cite{Horsley:1985dz,Horsley:1985fr}.
\begin{eqnarray}
\eta &=& - \int d^3 x' \int_{\!\!\!-\infty}^t d t_1 e^{\epsilon(t_1-t)}
           \int_{\!\!\!-\infty}^{t_1} d t' 
           \langle T_{12}({\bm x},t) T_{12}({\bm x}',t')\rangle_{ret} 
           \nonumber \\
\frac{4}{3}\eta + \zeta &=& - \int d^3 x' \int_{\!\!\!-\infty}^t d t_1 
         e^{\epsilon(t_1-t)} \int_{\!\!\!-\infty}^{t_1} d t'
         \langle T_{11}({\bm x},t) T_{11}({\bm x}',t')\rangle_{ret}            
\end{eqnarray}

These real-time Green's functions are obtained from their lattice (Euclidean
time) counterparts through the spectral function $\sigma(\omega)$. Quenched
lattice results have been obtained for these quantities \cite{Nakamura:2004sy}. 
In terms of the spectral function,
$\eta = \pi \, \lim_{\omega \rightarrow 0} \, \frac{\sigma(\omega)}{\omega}$.

\begin{figure}[htb]
\epsfxsize=4in
\centerline{\epsffile{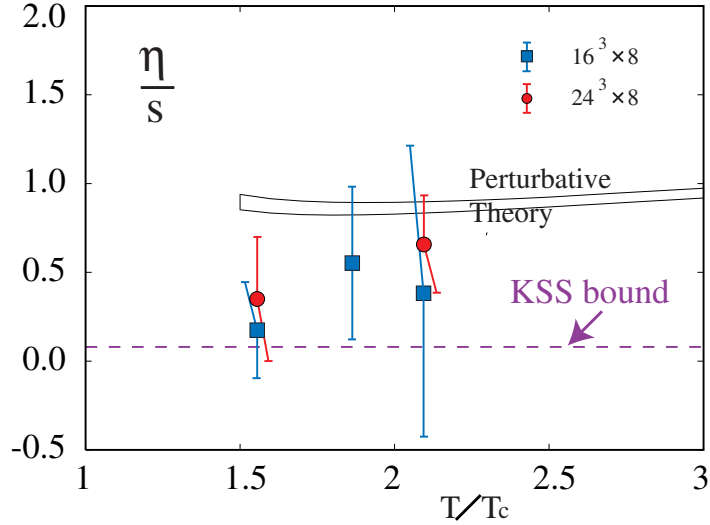}}
\caption{$\eta/s$ as a function of temperature. Dashed line is at $1/4\pi$
(from Ref.~\cite{Nakamura:2004sy}).}
\label{fig:nakamura}
\end{figure}

This means that one needs the spectral function near $\omega=0$, where it is
least well known. Thus the determination is difficult, and systematic as well
as statistical errors are large as shown in Fig.~\ref{fig:nakamura}.
The same paper finds $\zeta \approx 0$. Clearly there is a long way to go.

More recently, improved methods have been used to calculate the electrical
conductivity of the quark-gluon plasma \cite{Gupta:2003zh}.

\section{Lattice QCD at finite baryon number density}

Finite baryon/quark-number density is implemented by introduction of a 
quark-number chemical potential $\mu$. On the lattice this is achieved by
multiplying each of the links in the $+t$ direction by $e^\mu$ and each of 
those in the $-t$ direction by $e^{-\mu}$ in the quark action $S_f$.

Integrating out the fermion fields gives the determinant of the Dirac operator
which is complex, with a real part of indefinite sign. Standard simulation
methods, which rely on importance sampling, fail for such systems.

Some progress has been made in circumventing these problems for small $\mu$s
close to the finite temperature transition. Methods for doing this fall into 
several classes.
\begin{itemize}
\item
Analytic continuation: In the simplest case people simulate at imaginary $\mu$s,
where the fermion determinant is real and positive. The results are fitted to
a power series in $\mu^2$, which allows continuation to real $\mu$ 
\cite{deForcrand:2006hh,D'Elia:2004at}.
Fancier analytic continuation methods have also been used
\cite{Azcoiti:2005tv}.
\item
Power series expansions: These are similar in spirit to the analytic 
continuation methods. The exponential of the action is expanded in powers of
$\mu^2$. The coefficients of this expansion are then observables whose 
expectation values are measured in $\mu=0$ simulations
\cite{Ejiri:2003dc,Gavai:2004sd}.
\item
Reweighting methods: These start from a quark action with a positive fermion 
determinant, and reweight measurements by the ratio of the original fermion
determinant to this positive fermion determinant. One then divides by the
expectation value of this ratio of determinants \cite{Fodor:2004nz}.
\item
Phase quenched methods: One simulates using the magnitude of the fermion
determinant ignoring the phase. For small enough $\mu$ on a finite lattice,
the phase is small enough that this should yield the same phase structure as
the full simulation \cite{Sinclair:2006zm}.
\item
Canonical ensemble methods: The fermion determinant is projected on to states of
fixed quark number. One deals with any sign problems by reweighting.
\cite{Kratochvila:2005mk}
\end{itemize}

For 3-flavour QCD, it has been argued that the critical point observed at
zero chemical potentials, where the transition weakens from a first-order
phase transition to a crossover, would move to larger quark masses as $\mu$
is increased from zero, becoming the sort-after critical endpoint. Recent work
using analytic continuation from imaginary $\mu$ (de Forcrand and Philipsen
\cite{deForcrand:2006hh})
and simulations in the phase-quenched theory (Kogut and Sinclair)
\cite{Sinclair:2006zm}, indicate
that this does not happen. Instead, the critical mass appears to decrease with
increasing $\mu$.

Fodor and Katz, using the reweighting method, claim to find the critical 
endpoint, for physical quark masses \cite{Fodor:2004nz}. 
Their estimate of its position is
$T=162(2)$~MeV and $\mu=120(13)$~MeV. Since $\mu > m_\pi/2$, this is beyond
the reach of analytic continuation, phase quenched and series expansion methods.
It remains to develop other methods which could check this.
\begin{figure}[htb]
\epsfxsize=3.5in
\centerline{\epsffile{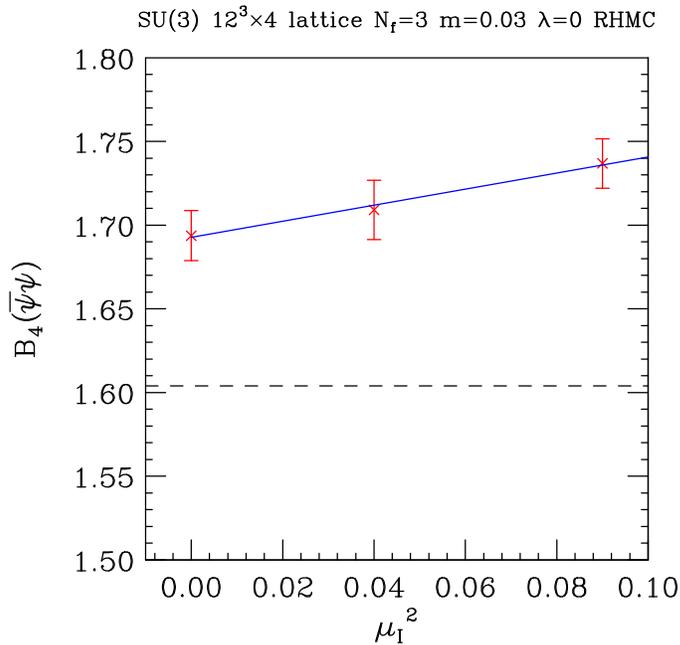}}
\caption{Binder cumulant for the chiral condensate as a function of $\mu_I$
(from Ref.~\cite{Sinclair:2006zm}).}
\label{fig:binder}
\end{figure}
Fig.~\ref{fig:binder} shows the behaviour of the Binder cumulant in the
phase-quenched theory. If there were a critical endpoint at (small) finite
$\mu_I$, this Binder cumulant would decrease through the Ising value
$1.604(1)$ (dashed line). Instead it increases indicating that there is no
critical endpoint in this range of $\mu_I$.

\section{The EOS at non-zero $T$ and $\mu$}

At finite temperature and small $\mu$, the series expansions in terms of $\mu$
also enable one to calculate such quantities as $p$, $s$ and $\epsilon$ and
hence to study the equation-of-state in terms of $T$ {\it and} $\mu$
\cite{Ejiri:2005uv}. Similar results have been obtained using multi-parameter
reweighting techniques \cite{Csikor:2004ik}.

\begin{figure}[ht]
\epsfxsize=4in
\centerline{\epsffile{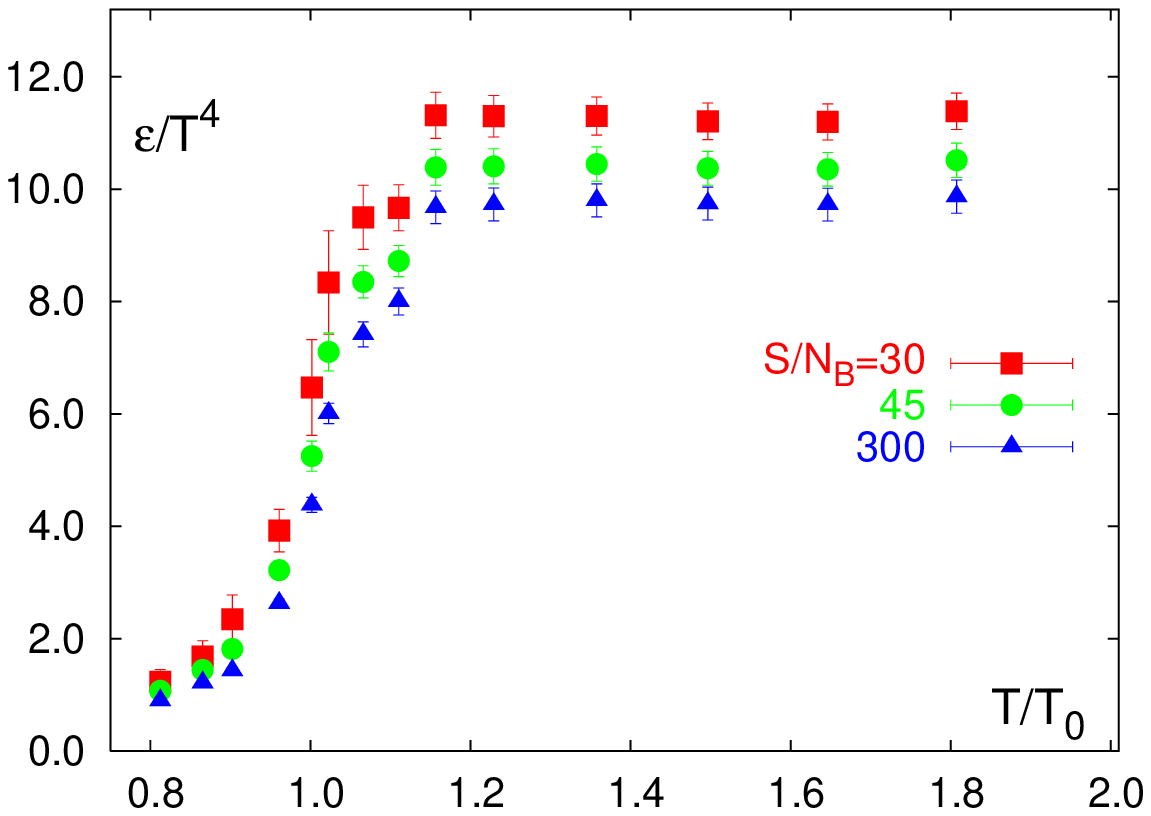}}
\centerline{\epsffile{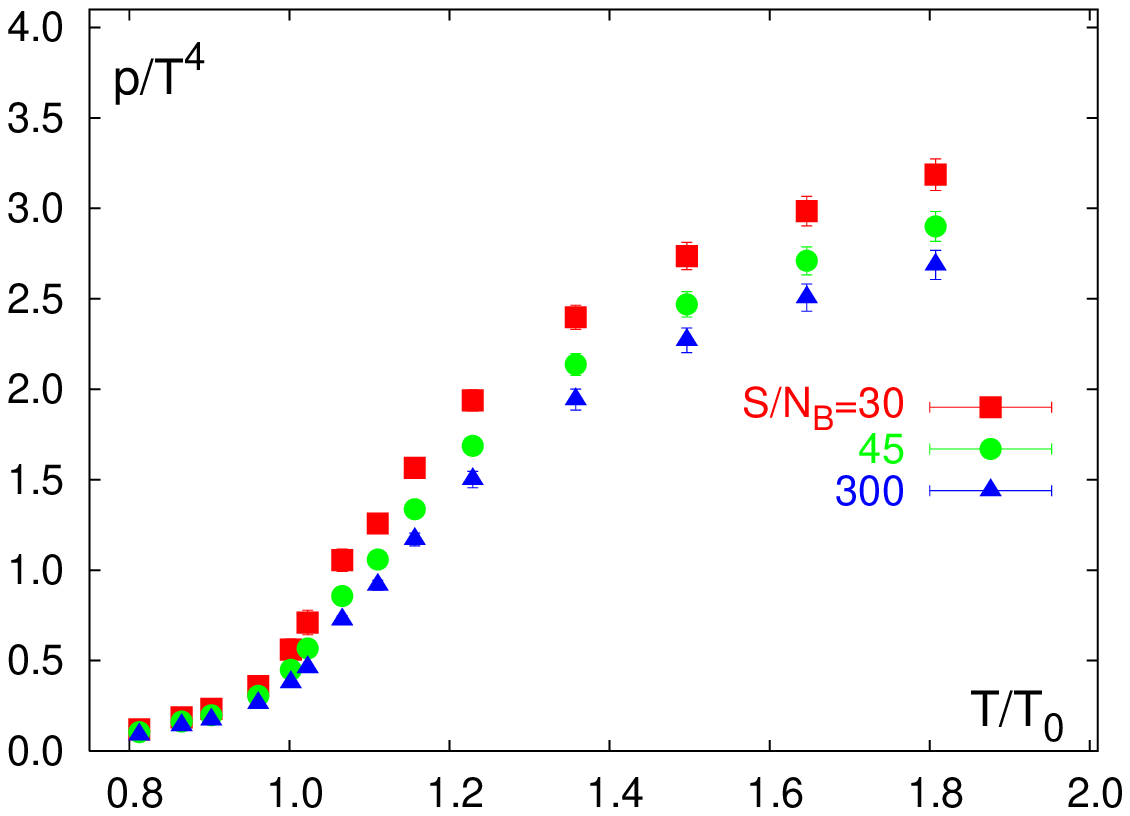}}
\caption{$\epsilon$ and $p$ as functions of T (from Ref.~\cite{Ejiri:2005uv}).}
\label{fig:bielefeld}
\end{figure}

The pressure $p$ can be obtained from measurement of the quark-number density
$\rho$, since
\begin{equation}
\rho = {\partial p \over \partial \mu}. 
\end{equation}
The pressure at zero $\mu$ is calculated as above. That at finite $\mu$ can be
obtained by integrating the previous equation. $\epsilon$ can also be 
calculated, but requires knowledge of the running of the coupling constant and
quark mass. Results from the Bielefeld group are shown in 
Fig.~\ref{fig:bielefeld}

\section{Summary and Conclusions}

Lattice QCD at finite temperature can probe the nature of the phase transition
from hadronic matter to a quark-gluon plasma. For physical quark masses this
appears to be a crossover without a true phase transition, influenced by the
second-order transition at $m_u=m_d=0$.

The QCD equation-of-state and other equilibrium thermodynamic quantities
measured in lattice QCD simulations provide input for an understanding of the
non-equilibrium thermodynamics at RHIC and the LHC.
Finite temperature (lattice) QCD probes QCD dynamics such as confinement and
chiral-symmetry breaking.

Charmonium spectral functions show that the $J/\psi$ remains intact at the
finite temperature transition $T_c$, but dissociates below $2 T_c$. This
should produce a reduction in $J/\psi$'s from the most energetic processes at
RHIC. Bottomonium spectral functions should show similar behaviour at the
even higher temperatures of the LHC heavy-ion program.

Just above $T_c$, the `deconfined' phase is a strongly-interacting fluid with
low viscosity.
Transport coefficients (including viscosity) need to be calculated as input
for hydrodynamic models. Early attempts have been made in lattice QCD 
simulations, but such measurements are difficult.

Sign problems hamper simulations at finite baryon/quark-number density. Some
progress has been made for small $\mu$s close to the finite temperature 
transition. 
The equation-of-state has been determined in this high-temperature low
baryon-number-density regime.
More work is needed to observe the critical endpoint, which is
the most striking feature expected in this region of the QCD phase diagram.
This is a region accessible experimentally to lower-energy relativistic
heavy-ion collisions.

New methods will be needed if one is ever to reach the high baryon-number
densities needed to understand the physics of neutron stars. At such densities
one expects such exotic states of matter as colour superconductors.

\section*{Acknowledgements}

Research supported by U.S. Department of Energy contract DE-AC-02-06CH11357.
I thank the authors of the manuscripts, from which some of the figures were
reproduced, for their permission to incorporate these figures in this
manuscript. My own research reported in this talk was performed in
collaboration with J.~B.~Kogut.

\end{document}